\documentstyle[epsf,aps]{revtex}
\begin{document}
\twocolumn[\hsize\textwidth\columnwidth\hsize\csname
@twocolumnfalse\endcsname

\title{Schwinger-boson approach to quantum spin systems: \\
Gaussian fluctuations in the ``natural'' gauge} 
\author{A. E. Trumper, L. O. Manuel, C. J. Gazza,\\ 
and H. A. Ceccatto}
\address{Instituto de F\'{\i}sica Rosario, Consejo Nacional de Investigaciones
Cient\'{\i}ficas y T\'ecnicas \\ and Universidad Nacional de Rosario, Bvd. 27
de Febrero 210 Bis, 2000 Rosario, Rep\'ublica Argentina }
\maketitle

\begin{abstract}

We compute the Gaussian-fluctuation corrections to the saddle-point
Schwinger-boson results using collective coordinate methods. Concrete
application to investigate the frustrated $J_1\!-\!J_2$ antiferromagnet on the
square lattice shows that, unlike the saddle-point predictions, there is a
quantum nonmagnetic phase for $0.53\! \lesssim \! J_2/J_1 \lesssim 0.64$. This 
result is
obtained by considering the corrections to the spin stiffness on large
lattices and extrapolating to the thermodynamic limit, which avoids the
infinite-lattice infrared divergencies associated to Bose condensation. The
very good agreement of our results with exact numerical values on finite
clusters lends support to the calculational scheme employed. 

\end{abstract} 


]

\narrowtext

In the last years there has been a lot of interest in the properties of
quantum magnetic systems,\cite{Manou} particularly frustrated quantum 
antiferromagnets. Although this interest was initially related to the possible
connections between magnetism and superconductivity in the ceramic compounds,
the current activity in the area is now well beyond this
original motivation.

Among the analytical methods used to study quantum spin systems, the
Schwinger-boson approach\cite{AA} is one of the most
elegant and successful techniques. Contrary to standard spin-wave
theory, it does not rely on having a magnetized ground-state, which leads to
nice rotational properties of the results and to the possibility of
describing ordered and disordered phases in an unified treatment. However,
this theory has the drawback of being defined in a constrained bosonic
space, with unphysical configurations being allowed when this
constraint is treated as a soft (average) restriction. This drawback can
be in principle corrected by including local fluctuations of the boson
chemical potential.\cite{RA}

Despite the widespread use in the literature of the Schwinger-boson
representation of quantum spin operators, we are not aware of a complete
calculation of Gaussian corrections to saddle-point results. In particular, 
for
frustrated quantum antiferromagnets such calculations have been sketched 
several times,\cite
{AA,S,A} but never fully undertaken. In this work we fill up
this gap by presenting the general calculation of Gaussian fluctuations in
the Schwinger-boson approach. Since the theory presents a local $U(1)$
symmetry, we use collective coordinate methods ---as developed in the
context of relativistic lattice gauge theories\cite{P}--- to handle the
infinitely-many zero modes associated to the local symmetry breaking
in the saddle-point expansion. As a concrete application, we study
the existence and location of the nonmagnetic quantum phase predicted to 
occur as a consequence of quantum fluctuations and frustration in
the $J_1\!-\!J_2$ model.\cite{CD}

We will consider a general Heisenberg Hamiltonian, $H=\sum_{\langle
ij\rangle }J_{ij}{\vec S_i}.{\vec S_j},$ where $\langle ij\rangle $ are
links on a lattice. We write spin operators in terms of Schwinger bosons:
\cite{A} ${\vec S_i}\!=\!{\frac 12}{\bf a}_i^{\dagger }.{\bf \vec \sigma }.
{\bf a
}_i$, where ${\bf a_i^{\dagger }}\!=\!(a_{i\uparrow }^{\dagger },a_{i\downarrow
}^{\dagger })$ is a bosonic spinor, ${\bf \vec \sigma }$ is the vector of
Pauli matrices, and there is a boson-number restriction $\sum_\sigma
a_{i\sigma }^{\dagger }a_{i\sigma }\!=\!2S$ on each site. With this
faithful representation of the spin algebra, the rotational 
invariant spin-spin
interaction can be written as  ${\vec S_i}.{\vec S_j}\!= :\!B_{ij}^{\dagger
}B_{ij}\!:-A_{ij}^{\dagger }A_{ij}$. We defined the $SU(2)$
invariants $A_{ij}=\frac 12\sum_\sigma \sigma a_{i\sigma }a_{j{\bar \sigma 
}}$ and  $B_{ij}^{\dagger }=\frac 12\sum_\sigma a_{i\sigma }^{\dagger
}a_{j\sigma }\ ({\bar \sigma }=-\sigma ,\ \sigma =\pm )$, and the notation
$:\!O\!:$ indicates the normal order of operator $O$.

Using boson coherent states to formulate the partition function,
we formally integrate the Schwinger bosons by decoupling the quartic
terms in $H$ using two complex Hubbard-Stratonovich fields 
$W_{ij}^r(\tau )$\ ($r=1,2$ correspond, respectively, to the fields that 
couple to the ferromagnetic $B$ and antiferromagnetic
$A$ channels). This leads to 
$Z\!=\!\int [{\cal D}\overline{W}
{\cal D}W][{\cal D}\lambda ]e^{-{\cal A}}$, where the measure 
$[{\cal D}\overline{W}{\cal D}
W]\!=\! \prod_{\langle ij\rangle ,r,\tau }(J_{ij}/2\pi i) d\overline{W}
^r_{ij}(\tau ) dW_{ij}^r(\tau)$ and $[{\cal D}\lambda]=\prod_{i,\tau}
d\lambda _i(\tau )/2\pi$. Here ${\overline W}$ is the complex conjugate of
$W$ and the effective action 
\begin{eqnarray}
\label{A}
{\cal A}=\int_0^\beta d\tau \left( \sum_{\langle ij\rangle ,r}J_{ij}
\overline{W}^r_{ij}(\tau )W_{ij}^r(\tau ) 
 +i2S\sum_i\lambda _i(\tau )\right) \nonumber \\
+{\cal F}_{\text{b}}(\overline{W},W,\lambda ) .\qquad \qquad \qquad 
\end{eqnarray}
The integration on $\lambda $ comes from the
integral representation of the $\delta $-functions which force the boson
number restriction on each site. 
In (\ref{A}) ${\cal F}_{{\rm b}}(\overline{W},W,\lambda )$ is the free
energy of a boson gas coupled to $\overline{W},W$. In a compact notation
where site ($i$), Nambu ($s$) and Matsubara ($\tau $) indices are 
summed over in the trace (Tr) operation, we 
have ${\cal F}_{\rm b}(\overline W,W,\lambda)\!=
\!{{\rm Tr}\ln}
{\cal M}(\overline W,W,\lambda )$. The
dynamical matrix ${\cal M}^{s s^{\prime}}_{ij}(\tau)$ is given by 
\begin{eqnarray}
{\cal M}_{ij}^{11}&=&\partial _\tau +i\lambda _i\delta _{ij}-{
\frac{J_{ij}}2}\left( W_{ij}^1-\overline{W}^1_{ji} \right) =
- {\overline {\cal M}}^{22}_{ij}, \nonumber \\
{\cal M}_{ij}^{12}&=&{\frac{J_{ij}}2}\left( W_{ij}^2-W_{ji}^2 
\right) =\overline{{\cal M}}^{21}_{ij}, \nonumber 
\end{eqnarray}
where we have suppressed the $\tau$ dependences to simplify
notation. It is important to stress that this formal result for 
${\cal F}_{\rm b}$ must be understood as the limit of a discrete-$\tau $
mesh.\cite{AS} In the functional formulation such a procedure picks up the
zero-point contributions to the energy coming from the Bogoliubov
transformation in the canonical approach.
 
The action ${\cal A}$ is invariant under both
local gauge transformations, $W^r_{ij} \rightarrow 
W^r_{ij} {\rm e}^{i(\theta_i \pm \theta_j)}$\ (-- for $r=1$, + for $r=2$), 
$\lambda _i \rightarrow \lambda _i-\partial _\tau \theta _i$, and global 
$SU(2)$ rotations of the original bosonic spinors.
Consequently, it would be
convenient to include external sources ${\bar \eta },\eta$ which explicitly
break the global $SU(2)$ invariance: ${\cal F}_{\rm b} \rightarrow
{\cal F}_{\rm b}-{\overline \eta} {\cal M}^{-1}{\eta}$. They are useful to 
study the
spontaneous breaking of this symmetry at $T=0$ as a consequence of Bose
condensation, which signals the onset of magnetic long-range order in the
ground state.\cite{A} However, in order to avoid the infrared divergencies
associated to this symmetry breaking, in what follows we approach the
thermodynamic limit by extrapolation of results on large but finite lattices.

Up to this point all the manipulations are exact. To proceed further we
resort, as usual, to a saddle-point expansion. Using the collective
notation for the fields 
${\vec \varphi}^\dagger \equiv (\overline{W},W,\lambda)$,
the (static) saddle-point
values $\overline{W}^r_{ij,0}, W_{ij,0}^r$ and $\lambda _{i,0}$ are obtained
from the extremal equations 
\begin{equation}
\label{EE}
\frac{\partial {\cal A}}{\partial {\vec \varphi}} \Big|_{\rm SP} =
{\vec \psi}_0^\dagger +{\rm Tr}\left( {\cal G}_0\frac {\partial {\cal M}_0}
{\partial {\vec \varphi}_0} \right) =0 , 
\end{equation}
where ${\vec \psi}_0^\dagger=(J\overline{W}_0,JW_0,2Si)$ and 
${\bf {\cal G}_0}\!=\!{\bf {\cal M}_0}^{-1}$ is the bosonic Green function at
saddle-point order.\cite{nota} After expanding 
${\cal A}$ to second order in the field fluctuations around these 
saddle-point values, we end up with 
\begin{equation}
\label{Z}Z\simeq {\rm e}^{-{\cal A}_0}\int [{\cal D}{\vec \varphi} ^{\dagger }
{\cal D}{\vec \varphi}]e^{-\frac 12 \Delta{\vec \varphi }^{\dagger }.
{\cal A}^{(2)}.\Delta {\vec \varphi}}.
\end{equation}
In this equation ${\cal A}_0\!=\!\beta \sum_{\langle ij \rangle}J_{ij} \left( 
|W_{ij,0}^1|^2-|W_{ij,0}^2|^2 \right)$ is the 
effective action 
evaluated at the saddle point, and $\Delta{\vec \varphi }$ 
are the field fluctuations. The matrix ${\cal A}^{(2)}$ 
is given by 
$$
{\cal A}^{(2)}\equiv
\frac{\partial ^2{\cal A}}{\partial {\vec \varphi}^{\dagger} \partial 
{\vec \varphi} } \Big|_{\rm SP}={\cal J}-{\rm Tr}\left( {\bf {\cal G}_0}
\frac{\partial 
{\bf {\cal M}_0}}{\partial {\vec \varphi}^{\dagger}_0 }{\bf {\cal G}_0}
\frac{\partial {\bf {\cal M}_0}}{\partial {\vec \varphi}_0}\right),
$$
where ${\cal J}$ is a diagonal matrix containing the exchange couplings
$J_{ij}$ along the diagonal, except for the entries corresponding
to $\lambda\!-\!\lambda$ derivatives that are zero.

We stress that the saddle-point expansion of the effective action breaks the
gauge invariance of the theory. Although the spontaneous breaking of a local
symmetry is strictly forbidden by Elitzur's theorem,\cite{E} a possible
justification of this procedure has been given in the context of
relativistic lattice gauge theories:\cite{BD} The existence of
saddle-point solutions connected by the continuous $U(1)$
gauge group makes the ${\cal A}^{(2)}$ matrix to have infinitely-many 
zero modes, which
are the Goldstone bosons associated to the spurious local symmetry breaking. 
In particular, for
translational-invariant saddle-point values $W_{ij,0}^r\!=\!W_{i-j,0}^r,\
\lambda _{i,0}\!=\!\lambda _0$, transforming to momentum-frequency space there
is a zero mode ${\vec \varphi }_0^R({\bf k},\omega )\!=\!{\delta 
{\vec \varphi }
(\theta )/{\delta \theta _{{\bf k}\omega }}|_{{\rm SP}}\ }$ in every ${\bf k}
\!-\!\omega $ subspace. Here ${\vec \varphi }(\theta )$ is the vector of
gauge-transformed fields (Notice that ${\vec \phi }_0^R$ is a
{\it right} eigenvector of the nonhermitian matrix ${\cal A}^{(2)}$). To 
avoid the
infinities associated to these modes without restoring forces ---which
correspond to local symmetry transformations--- we introduce collective
coordinates along the gauge orbit. Exact integration of these coordinates
eliminates the zero modes and restore the gauge symmetry (in the sense
that noninvariant operators average to zero).\cite{BD} This program can be 
carried out by enforcing in 
(\ref{Z}) the so-called background gauge condition or ``natural'' gauge,
\cite{P} ${\vec \varphi }(\theta )^{\dagger }.{\vec \varphi }_0^R=0$.  
This condition can be introduced into the functional measure by means of
the Faddeev-Popov trick, and restricts the integration to fields 
fluctuations which are orthogonal to the collective coordinates. At 
$T=0$, after carrying out the
remaining integrations in (\ref{Z}) on the genuine fluctuations, we
obtain the one-loop correction to the ground-state energy {\it per site}, 
\begin{equation}
\label{E1}
E_1=-\frac 1{2\pi }\int_{-\infty }^\infty d\omega \sum_{{\bf k}%
}\ln \left( \frac{\Delta _{{\rm FP}}({\bf k},\omega )}{|\omega| \sqrt{\det 
{\cal A}^{(2)}_{\perp }({\bf k},\omega )}}\right) .
\end{equation}
Here the Fadeev-Popov determinant $\Delta _{{\rm FP}}({{\bf k},\omega })\!
=\!\left| 
{\vec \varphi }_0^L({{\bf k},\omega }).{\vec \varphi }_0^R({{\bf k},\omega }
)\right| $ (${\vec \varphi }_0^L({{\bf k},\omega })$ is the {\it left}
zero mode of ${\cal A}^{(2)}$ in the ${\bf k}\!-\!\omega$ subspace), and 
${\cal A}^{(2)}_{\perp }$ is the projection of 
${\cal A}^{(2)}$ 
in the subspace orthogonal to the collective coordinates. 

We have particularized these results for a matrix $J_{ij}$ which couples a
given site to its first ($J_1$) and second ($J_2$) neighbors on a square
lattice. This model with frustrating $J_1,J_2>0$ has been much studied in
the last years in connection with the physics of the lightly doped CuO$_2$
planes (see \cite{DS} and references therein). The main problems under 
discussion are the location and
physical properties of a possible nonclassical phase for intermediate
frustration $\alpha\!=\!J_2/J_1\!\sim\!1/2$. In the ordered phases of this 
model
(with magnetic wavevectors ${\bf Q}\!=\!(\pi ,\pi )$ for small $\alpha $ 
and ${\bf Q}\!=\!(\pi ,0)$ for $\alpha\!\sim\!1$) the Bose condensate breaks 
the global $SU(2)$ symmetry and its density gives the local 
magnetization.\cite{A} The
associated physical Goldstone modes at ${\bf k}\!=\!0,{\bf Q}$\ lead to
serious infrared divergencies of intermediate quantities, which have to be
cured by standard renormalization prescriptions. In order to avoid
these problems we have computed physical quantities (which are free of these
divergencies) on large by finite lattices, and finally extrapolated these
values to the thermodynamic limit. However, this procedure requires a
careful numerical treatment, since finite results are obtained by 
cancellations among products of the very large entries of the matrix 
${\cal A}^{(2)}$.

The equations (\ref{EE})-(\ref{E1}) lead to a ground-state energy 
$E_{\rm GS}\!=\!E_0+E_1$ as shown in Fig. 1. This figure
contains the result for the infinite lattice and also for finite lattices 
of 16 and 36 sites, which allow a comparison
with exact results obtained by numerical diagonalization.\cite{Sch} The
addition of the Gaussian correction $E_1$ (eq. (\ref{E1})) slightly improves
the already very good saddle-point value $E_0$. At saddle-point order the 
theory
predicts a first order transition between the two magnetic ground states at
some intermediate frustration ($\alpha \simeq 0.6$), with no intervening
disordered phase. This is in contradiction with numerical studies\cite{ES}
and series expansion results,\cite{G} which found a disordered 
purely-quantum phase somewhere in the range $0.3\lesssim\! \alpha\!
\lesssim 0.6$. 
However, it
has been recently shown that these finite-lattice studies probably did not
reach the scaling region where the required extrapolation can be trusted.
\cite{FGTC} In our case, the existence or not of magnetic long-range order 
was investigated by considering the spin-stiffness tensor 
$\rho_{\rm s}=\partial ^2E({\bf Q})/\partial{\bf Q}
\partial {\bf Q}$, where $E_{\rm GS}({\bf Q})$ is the ground-state energy with 
twisted boundary
conditions. \cite{ES,FGTC} In particular, the corrections to the saddle-point
values obtained 
in \cite{FGTC} lead to the results shown in Fig. 2. It gives the stiffness 
along one of 
the lattice directions in the ${\bf Q}=(\pi ,\pi )$ phase at saddle-point and 
one-loop order, on lattices of 16
and 20 sites (according to numerical studies\cite{ES} the tilted 20-site
lattice has a stiffness closer to the thermodynamic limit than  
regular clusters). The comparison with exact results\cite{Sch} shows a 
good
agreement, especially for the 20-site lattice. The extrapolation to the
thermodynamic limit leaves a window $0.53\lesssim\! \alpha\! 
\lesssim 0.61$ where
the stiffness vanishes and the magnetic order is melted by the combined
action of quantum fluctuations and frustration (see Fig. 3). This 
result is in fairly good agreement with the second-order modified
spin wave calculation of \cite{G}, which predicts a quantum disordered phase
in the range $(0.52,0.57)$. Notice, however, that
our theory --valid also in the disordered region-- predicts
that for $0.61\lesssim\! \alpha\! \lesssim 0.64$ the short-range
antiferromagnetic order is energetically more favourable than the weak
collinear order (see dotted line in Fig. 1). Finally, we have computed
the spin-wave velocity $c_{\rm s}$ in the N\'eel phase using the 
finite-size formula
$E_{\rm GS}(N)\!\sim\!E_{\rm GS}-a/N^{3/2}$ and the relation
$a\!=\!1.4372c_{\rm s}$.\cite{NZ} For 
$J_2\!=\!0$ we obtained $c_{\rm s}\!\simeq\!1.52J_1$ at saddle-point order, 
and
$c_{\rm s}\!\simeq\!1.37J_1$ including the Gaussian fluctuations. These values
should be compared to the result of exact diagonalization,\cite{Sch} 
$c_{\rm s}\! \simeq\! 1.28-1.44J_1$, which depends on the cluster sizes 
considered 
to extrapolate. In Fig. 3 we plot the scaling coefficient $a$ as a function
of frustration (in the collinear phase this coefficient should be 
proportional to
some anisotropy-averaged velocity). In the N\'eel phase the spin-wave
velocity vanishes at $\alpha \!\simeq\! 0.54$, slightly above the
point where the stiffness goes to zero. In the
collinear phase the averaged velocity never vanishes, and actually it blows 
up at $\alpha\! \simeq\!0.58$. Notably, this is the value where the
stronger stiffness in the antiferromagnetic direction of the collinear order
vanishes. More details and a further discussion of these results will be 
given elsewhere.    

Before closing, it is worth mentioning a few side questions we have
considered while performing these calculations. First, since the Hamiltonian
commutes with the local boson-number restrictions, we considered imposing
them only through $\tau $-independent Lagrange multipliers $\lambda _i$.
This reduces the local gauge symmetry to static transformations and,
consequently, ${\cal A}^{(2)}$ contains zero modes only in the $\omega\!=
\!0$ subspace. For the calculation of the ground-state energy ---which involves
an integration over $\omega $--- it is then allowed to simply forget about
these modes and compute $E_1\!=\!\frac 1{4\pi }\int_{-\infty }^\infty d\omega
\sum_{{\bf k}}\ln \det {\cal A}^{(2)}_{{\rm trunc}}({\bf k},\omega )$, where 
${\cal A}^{(2)}_{{\rm trunc}}$ is equal to ${\cal A}^{(2)}$ with the column 
and row
corresponding to the (static) $\lambda $ fluctuations deleted. In this case 
there is no
need for $\Delta _{{\rm FP}}$ and (\ref{E1}) is equivalent to the RPA result
for the bosonic theory with $\lambda _0$ taken as a chemical potential. We
checked that this procedure gives the same result that (\ref{E1}), although
the numerical evaluation requires extra care because of the divergencies
near $\omega\!=\!0$. Second, in most works in the literature on 
Schwinger bosons
the identity $:\!B_{ij}^{\dagger }B_{ij}\!:\negthinspace
+ A_{ij}^{\dagger }A_{ij}\!\equiv\! S^2$ (which holds because of the 
constraint)
is used to simplify the Hamiltonian, leaving only the ferromagnetic $B$ or
the antiferromagnetic $A$ channel in the formulation. In a previous
publication\cite{CGT} we warned that at saddle-point order this produces
large errors in the ground-state energy, since the above identity is largely
violated when the constraint is imposed only on average. We proved that this
remains true even after the inclusion of fluctuations, with the
contributions of both channels being important. 
Third, in (\ref{E1}) we can write\  $\ln \det 
{\cal A}^{(2)}_\perp\!\equiv\!{\rm Tr} \ln {\cal A}^{(2)}_{\perp}\!=\!
- \sum_{n=1}^\infty {\frac 1n}
{\rm Tr} ({\it I} - {\cal A}^{(2)}_{\perp})^n$. If only the first 
term $(n\!=\!1)$ is added to the saddle-point energy $E_0$, the result is 
equivalent
to the energy obtained in a full Hartree-Fock-Bogoliubov decoupling of the 
quartic boson interactions
in $H$. Moreover, it can be proved that $E_{\rm HFB}\!=\!E_0+E_{n=1}\!
=\!{\frac 32}
E_0$. That is, a fully self-consistent treatment takes advantage of the
unphysical enlargement of Fock space to lower the energy too much. The 
inclusion of terms with $n\geq 2$ corrects $E_{\rm HFB}$ in nearly
50\%. 
   
In conclusion, we have computed the Gaussian-fluctuation corrections to the
Schwinger-boson saddle-point results using collective coordinate methods. As
a concrete application, we investigated the ground-state structure of the $
J_1\!-\!J_2$ model. By considering the spin stiffness of this model we showed
that, contrary to the saddle-point predictions, there is a quantum
nonmagnetic phase that intervenes between the N\'eel $(\pi,\pi)$ and
collinear $(\pi,0)$ orders, as suggested by numerical methods. Its stability
region is, however, smaller than predicted by these methods. Moreover, the
comparison with exact results on finite lattices lends support to our
calculations, which, in addition, have the advantage of a well-defined
thermodynamic limit. Finally, the theory developed here can be easily extended
to spiral phases,\cite{CGT} which would allow to investigate, for instance, 
the ground-state order in triangular\cite{GC} and {\it kagom\'e} lattices.
\cite{MGTC} Work in this direction is in progress.

\figure{FIG. 1: Ground-state energy {\it per site} $E_{\rm GS}$ as a 
function of $J_2/J_1$ 
for lattices of (from top to bottom) $N=\infty$, 36, and 16 sites. The full 
and dashed lines give the fluctuation-corrected and saddle-point results, 
respectively. Points are exact results from \cite{Sch}. The dotted line
in the range $0.53 \lesssim J_2/J_1 \lesssim 0.64$ indicates the region 
without long-range magnetic order. Notice that for clarity the 
curves corresponding to $N=\infty$ and 36 have been shifted upwards in 
0.1 and 0.05 respectively.}

\figure{FIG. 2: Spin stiffness $\rho_{\rm s}$ in the N\'eel phase for a) a
16-site lattice, and b) a tilted 20-site lattice. Symbols and
linetypes are the same as in Fig. 1.}

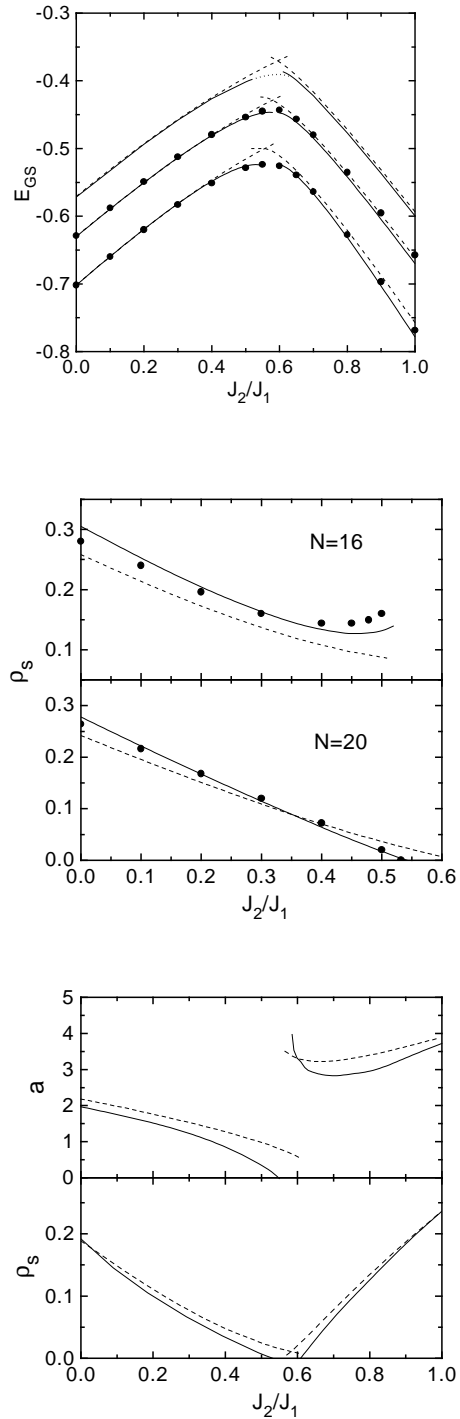
\figure{FIG. 3:  Scaling coefficient $a$ proportional to the spin-wave 
velocity (top panel) and spin stiffness $\rho_{\rm s}$ (bottom panel)
extrapolated to the thermodynamic limit. Linetypes are the same as in
Fig. 1.} 

\begin{figure}[ht]
\vspace{-0.25cm}
\hspace{1cm}
\epsfysize=6cm
\leavevmode
\epsffile{fig1.eps}
\end{figure}

\begin{figure}[h]
\vspace{-0.25cm}
\hspace{1cm}
\epsfysize=6cm
\leavevmode
\epsffile{fig2.eps}
\end{figure}

\begin{figure}[h]
\vspace{-0.25cm}
\hspace{1cm}
\epsfysize=6cm
\epsffile{fig3.eps}
\end{figure}

\begin{thebibliography}{99}
\bibitem{nota} Notice that since the ferromagnetic channel $B$ contributes
with a positive sign to the spin-spin interaction in $H$, the solution to 
(\ref{EE}) corresponds to an analytic continuation in which $W^1_{ij,0}$ 
is real but $\overline{W}_{ij,0}^1\!=\!-W_{ij,0}^1$.
\end{thebibliography}
\end{document}